\documentclass[12pt]{article}
\usepackage{amsmath,amssymb,exscale}  
\input epsf

\def\gappeq{\mathrel{\rlap {\raise.5ex\hbox{$>$}}
{\lower.5ex\hbox{$\sim$}}}}
\def\permil{$\%\raise.20ex\hbox{$_0$}}
\def\lappeq{\mathrel{\rlap{\raise.5ex\hbox{$<$}}
{\lower.5ex\hbox{$\sim$}}}}
   
\newcommand{\un}{\underline}

\begin{document}
\topmargin -1.0cm
\oddsidemargin -0.8cm
\evensidemargin -0.8cm
\pagestyle{empty}
\begin{flushright}
UNIL-IPT-00-17\\
IC/2000/134\\
hep-th/0008079\\
August 2000
\end{flushright}
\vspace*{5mm}

\begin{center}

{\Large\bf Fermion zero-modes on brane-worlds \\}
\vspace{1.0cm}

{\large Seif Randjbar-Daemi$^a$  and 
Mikhail Shaposhnikov$^b$}\\
\vspace{.6cm}
{\it {$^{a}$International Center for Theoretical Physics, Trieste,
Italy}}\\
{\it {$^{b}$Institute of Theoretical Physics, University of Lausanne,
\\
CH-1015 Lausanne, Switzerland}}\\

\vspace{.4cm}
\end{center}

\vspace{1cm}
\begin{abstract}

We study localization of bulk fermions on a brane with inclusion of
Yang-Mills and scalar backgrounds in higher dimensions and give the
conditions under which localized chiral fermions can be obtained. 

\end{abstract}

\vfill

\eject
\pagestyle{empty}
\setcounter{page}{1}
\setcounter{footnote}{0}
\pagestyle{plain}


{\bf Introduction.} Suggestions that extra dimensions may not be
compact \cite{RS1}-\cite{rusu} or large \cite{ant,ant1} can provide
new insights for a solution of gauge hierarchy problem \cite{ant1},
of cosmological constant problem \cite{RS2,seif}, and give new
possibilities  for model building. One of the interesting questions,
related to these ideas, is localization of different fields on a
brane. It has been shown that the graviton \cite{rusu} and the
massless scalar field \cite{baj} have normalizable zero modes on
branes of different types, that the abelian vector fields are not
localized in the Randall-Sundrum (RS) model in five dimensions but
can be localized in some higher-dimensional generalizations of it
\cite{oda}. In contrast, in \cite{baj} it was shown that fermions do
not have normalizable zero modes in five dimensions, while in
\cite{oda} the same result was derived for a compactification on a
string \cite{gs,rg} in six dimensions\footnote{Bulk fermions were
localized on brane world studied some years ago in \cite{GW} by the
use of a bulk magnetic field, which falls out of the framework of our
ansatz. However, these solutions generally have problems with
normalizability of the graviton modes.}. It is known, though, that
fermion interaction with a scalar domain wall in five dimensions can
lead to localization of chiral fermions \cite{RS1}. Gauge field
localization by confinement effects were discussed in
\cite{Dvali:1997xe}, bulk fields in a slice of AdS in
\cite{Gherghetta:2000qt}.

In this note we shall prove that under quite general assumptions 
about the geometry and topology of the internal manifold of the
higher-dimensional warp factor compactification there exist massless
Dirac fermions. However, these fermionic modes are generically
non-normalizable. On the other hand if we include a Yukawa-type
coupling to a scalar field of a domain-wall type we can ensure
chirality as well as localization of the fermions. To generate chiral
fermions by this mechanism the topology of the internal Kaluza-Klein
manifold and the gauge field defined on it should be such that the
index of the Dirac operator defined on this manifold is non-zero. At
the end of this note we shall mention the example of the $K_3$
surface and $S^4$ with a background instanton configuration defined
on it.

{\bf Branes with gauge and gravity backgrounds.} We shall consider 
$D = D_1+D_2+1$ - dimensional manifolds with the geometry
\begin{equation}
\label{eq2}
{\rm d}s^2={\rm e}^{A(r)}\eta_{\mu\nu}{\rm d}x^\mu{\rm d}x^\nu
+{\rm e}^{B(r)}g_{mn}(y){\rm d}y^m{\rm d}y^n+{\rm d}r^2,
\end{equation}
where $\mu,\nu=0,1\dots,D_1-1, \quad m,n=1,\dots,D_2$. The coordinates $y^m$ 
cover an internal manifold $K$ with the metric $g_{mn}(y)$. The 
$D$-dimensional Dirac equation is
\begin{equation}\label{eq3}
\Gamma^A E_A^M(\partial_M-\Omega_M+A_M)\Psi(x,y,r)=0,
\end{equation}
where $E_A^M$ is the vielbein, $\Omega_M= \frac{1}{2}
\Omega_{M[AB]}\Sigma^{AB}$ is the spin connection,$\quad
\Sigma_{AB}=\frac{1}{4}[\Gamma_A,\Gamma_B]$, and $A_M$ is a
Yang-Mills field in the algebra of some  gauge group $G$. The RS
model is the special case with $D_2=0$ and $A_M=0$.

The  non-vanishing components of $\Omega_M$ are
\begin{eqnarray}
\Omega_\mu&=& \frac{1}{4}A'{\rm e}^{A\over 
2}\delta_\mu^a\Gamma_r\Gamma_a\label{eq4}~,\\
\Omega_m&=& \frac{1}{4}B'{\rm e}^{\frac{B}{2}}e_m^{\un{a}}
\Gamma_r\Gamma_{\un{a}}+\omega_m ~,
\label{eq5}
\end{eqnarray}
where $\Gamma_r,\Gamma_a,~a=0,1,\dots,D_1-1$ and
$\Gamma_{\un{a}},~\un{a}=1,\dots,D_2$  are the constant Dirac
matrices and $\omega_m=\frac{1}{8}
\omega_{m[\un{a},\un{b}]}[\Gamma_{\un{a}},\Gamma_{\un{b}}]$ is the
spin  connection derived from the metric
$g_{mn}(y)=e_m^{\un{a}}e_n^{\un{b}}\delta_{\un{a}\un{b}}$.

Assume $A_\mu=A_r=0$. The Dirac equation then becomes
\begin{equation}
\label{eq6}
\left\{{\rm e}^{-\frac{A}{2}}\not\hspace{-0.1cm}
{\partial}_x+\Gamma^r\left(\partial_r
+{D_1\over 
4}A'+{D_2\over 4}B'\right)+{\rm e}^{-{B\over 2}}
\not\hspace{-0.1cm}{\Delta}_y\right\}\Psi=0,
\end{equation}
where $\not\hspace{-0.1cm}{\Delta}_y=\Gamma^{\un{a}}{\rm
e}_{\un{a}}^m(\partial_m-\omega_m+A_m)$ is the  Dirac operator on the
internal manifold $K$ and in the background of the gauge  field
$A_m$. With an appropriate choice of $K$ and $A_m$ this operator will
have zero modes \cite{witten}. Denote these modes by $\psi(y)$. We
can then write
\begin{equation}
\label{eq7}
\Psi(x,y,r)=\psi(y)f(r)\phi(x)~,
\end{equation}
where $f$ and $\phi$ satisfy
\begin{eqnarray}
\label{eq8}
\not\hspace{-0.1cm}{\partial}_x\phi(x)&=&0~,\\
f'+\left({D_1\over 4}A'+{D_2\over 4}B'\right)f&=&0,
\label{eq9}
\end{eqnarray}
or
\begin{equation}
\label{eq10}
f(r)=\mbox{const. }{\rm e}^{-\left({D_1\over 4}A+{D_2\over 4}B\right)}.
\end{equation}
The effective Lagrangian for $\phi$ then becomes
\begin{equation}
\int{\rm d}r{\rm d}y\sqrt{-G}{\bar \Psi}\Gamma^A E_A^M(\partial_M-\Omega_M+A_M)\Psi=
\bar{\phi}(x)\not\hspace{-0.1cm}{\partial}_x\phi(x)\times
\int {\rm e}^{-{A\over 2}}
{\rm d}r {\rm d}y  \sqrt{g}\psi^{\dagger}(y)\psi(y).
\label{norm}
\end{equation}
This should be compared with the expression of $D_1$-dimensional
Newton constant $G_{D_1}$ in terms of the $D$-dimensional one,
\begin{equation}
        G_{D_1}^{-1} = G_{D}^{-1} V_{D_2} \int d\,r\,
	\exp{\left((\frac{D_1}{2}-1)A +\frac{B}{2}\right)}~,
\label{planck}
\end{equation}
where $V_{D_2}$ is the volume of the manifold $K$.
Thus, to have the localization of gravity and finite kinetic energy
for $\phi$, both integrals (\ref{norm},\ref{planck}) must be
simultaneously finite. This is not the case for the exponential
warp-factor $ A \propto - |r|$ considered in the literature so far. 
In fact, for such $A$ and $B$, the function $f$ in (\ref{eq10})
diverges as $r \to \infty$.

So, for presently known solutions, the bulk fermions cannot be
localized on a brane with the use of gravity and gauge fields only.

{\bf Yukawa Coupling and Chiral Fermions.}
Now let us include a real scalar field $\chi$ in our problem. The
modification of the  Dirac equation will be through some Yukawa term,
with the coupling $\lambda$, viz.
\begin{equation}\label{eq11}
\left\{{\rm e}^{-{A\over 
2}}\not\hspace{-0.1cm}{\partial}_x+\Gamma^r\left(\partial_r+{D_1\over
4}A'+{D_2\over  4}B'\right)+\lambda\chi+{\rm e}^{-{B\over
2}}\not\hspace{-0.1cm}{\Delta}_y\right\}\Psi=0.
\end{equation}
The details of the $\chi$-field dynamics will not be important for our 
discussion. We shall only assume that its equation of motion admits a 
localized $r$-dependent solution such that $\chi(r)\to|v|\epsilon(r)$ as 
$|r|\to\infty$, where $v=\langle\chi\rangle$, and $\epsilon(r)$ is the sign 
function. With this assumption and imposing the chirality condition
$\Gamma^r \Psi = + \Psi$, far away from the core region we need to solve
\begin{eqnarray}
\not\hspace{-0.1cm}{\partial}_x\phi&=&0, \nonumber \\
\left(\partial_r+{D_1\over 4}A'+{D_2\over 
4}B'+\lambda|v|\epsilon(r)\right)f&=&0, \nonumber\\
\not\hspace{-0.1cm}{\Delta}_y\psi&=&0.
\label{eq12}
\end{eqnarray}
The solution of the above equation can 
be written as
\begin{equation}\label{eq13}
\psi(x,y,r)={\rm e}^{-\left({D_1\over 4}A+{D_2\over 4}B\right)-
\lambda|v|r\epsilon(r)}\cdot \psi(y)\phi(x),
\end{equation}
where $\not\hspace{-0.1cm}{\Delta}_y\psi(y)=0$. 

Thus, to have localized $\Psi(x,y,r)$ it is sufficient that
\begin{equation}
\label{eq14}
-{A\over 2}-2\lambda|v|r\epsilon(r)<0.
\end{equation}
This can be achieved for large enough values of $\lambda |v|$. For
example, for solutions of Einstein equations with  $A = B = c r
\epsilon(r)$,  where $c < 0$, that can be obtained for a string in 6
dimensions \cite{gs} or on $K=K3$ in higher dimensions \cite{grs},
it is sufficient to have $\lambda |v|>-c/4$. 

Now we come to the issue of chirality of the normalizable zero modes.
First we note that for an even $D_2$ normalizable solutions of 
$\not\hspace{-0.2cm}{\Delta}_y\psi(y)=0$ have definite chirality. The
index theorem gives the difference $n_+-n_-$, where $n_+$ and $n_-$
are respectively the number of positive and negative chirality zero
modes of the operator $\not\hspace{-0.1cm}{\Delta}_y$. Since we have
imposed $\Gamma^r =1$, the chiralities of $\psi(y)$ and $\phi(x)$
will be identical\footnote{This is due to the fact that for an odd
$D$ $\Gamma^r = \Gamma_{D_1}\cdot\Gamma_{D_2}$ where $\Gamma_{D_1}$
and $\Gamma_{D_2}$ are the chirality matrices in $D_1$ and $D_2$
dimensions.}. Thus the number of chiral families will be equal to
$n_+-n_-$. This mechanism is identical to the one which generates
chiral fermions in the standard Kaluza-Klein compactification
\cite{seif1}.

On the example of a  $K=K_3$ compactification we obtain two chiral
families, while for $K=S^4$ with an SU(2) instanton on it there will
be  $\frac{2}{3} t(t+1)(2t+1)$ chiral localized families in 4
dimensions,  where t is the spin of the fermion representations. For
$D=7$ we can take $K=S^2$ with a U(1) magnetic monopole field of
charge $n$ on it. The number of chiral families will then be equal to
$n$ \cite{seif2}.

{\bf Conclusions.} We defined conditions under which bulk
fermions can be localized on a brane in models with localized gravity
in higher dimensional generalizations of the RS model if only gauge
and gravitational backgrounds are considered. We show how the domain-wall
scalar field structures can insure localization and chirality at the
same time. The number of chiral fermions is related to the topology
of the manifold $K$ and the gauge field background.

It remains to be seen if one can find solutions which incorporate all
the required features, namely, localized fields of various spins with
the correct standard model quantum numbers in a non-singular
background geometry. The non-singularity of the localized geometry
seems to be rather difficult to achieve, at least without precence of
a brane. It has been shown in \cite{seif} that for the metrics of the
type given in eq.(\ref{eq2}) which are regular at $r=0$ the vacuum
Einstein equations for $D_2>1$ produce generally singular solutions,
although with a finite volume in the $y,r$ subspace. It has been
recently argued by Witten \cite{Witt} that such naked singularities
make the physical interpretation of these solutions problematic.

\noindent {\it Acknowledgments:} We wish to thank T. Gherghetta
for helpful discussions. This work was supported by the
FNRS,  contract no. 21-55560.98.


\begin{thebibliography}{99}

\bibitem{RS1}
V.~A.~Rubakov and M.~E.~Shaposhnikov,
Phys.\ Lett.\  {\bf B125} (1983) 136.

\bibitem{RS2}
V.~A.~Rubakov and M.~E.~Shaposhnikov,
Phys.\ Lett.\  {\bf B125} (1983) 139.

\bibitem{akama}
K.~Akama, in {\it Proceedings of the Symposium on Gauge Theory and 
Gravitation}, Nara, Japan, eds. K.~Kikkawa, N.~Nakanishi and H.~Nariai 
(Springer-Verlag, 1983);\\
M.~Visser,
Phys.\ Lett.\  {\bf B159} (1985) 22.

\bibitem{seif}
S.~Randjbar-Daemi and C.~Wetterich,
Phys.\ Lett.\  {\bf B166} (1986) 65.

\bibitem{rusu}
L.~Randall and R.~Sundrum,
Phys.\ Rev.\ Lett.\  {\bf 83} (1999) 4690.

\bibitem{ant}
I.~Antoniadis,
Phys.\ Lett.\  {\bf B246} (1990) 377.

\bibitem{ant1}
N.~Arkani-Hamed, S.~Dimopoulos and G.~Dvali,
Phys.\ Lett.\  {\bf B429} (1998) 263;\\
I.~Antoniadis, N.~Arkani-Hamed, S.~Dimopoulos and G.~Dvali,
Phys.\ Lett.\  {\bf B436} (1998) 257.

\bibitem{baj}
B.~Bajc and G.~Gabadadze,
Phys.\ Lett.\  {\bf B474} (2000) 282.

\bibitem{oda}
I.~Oda,
hep-th/0006203.

\bibitem{gs}
T.~Gherghetta and M.~Shaposhnikov,
Phys.\ Rev.\ Lett.\  {\bf 85} (2000) 240.

\bibitem{rg}
R.~Gregory,
Phys.\ Rev.\ Lett.\  {\bf 84} (2000) 2564.

\bibitem{GW} 
G. W. Gibbons and D. L. Wiltshire, Nucl. \ Phys. \ {\bf B287} (1987)
717.

\bibitem{Dvali:1997xe}
G.~Dvali and M.~Shifman,
Phys.\ Lett.\  {\bf B396} (1997) 64.

\bibitem{Gherghetta:2000qt}
T.~Gherghetta and A.~Pomarol,
hep-ph/0003129.

\bibitem{witten}
E.~Witten, ``Fermion Quantum Numbers In Kaluza-Klein Theory,''
{\it  In T. Appelquist, (Ed.) et.al.: 
Modern Kaluza-Klein Theories, 438-511}. 

\bibitem{grs}
S.~Randjbar-Daemi and M.~Shaposhnikov,
hep-th/0008087.

\bibitem{seif1}
S.~Randjbar-Daemi, A.~Salam and J.~Strathdee,
Phys.\ Lett.\  {\bf B132} (1983) 56.

\bibitem{seif2}
S.~Randjbar-Daemi, A.~Salam and J.~Strathdee,
Nucl.\ Phys.\  {\bf B214} (1983) 491.

\bibitem{Witt}
E.~Witten,
hep-ph/0002297.

\end{thebibliography}
\end{document}